\documentclass [11pt,a4paper]{article}
\usepackage{amsthm,amssymb}
\usepackage[lined,algonl,boxed]{algorithm2e}
\newtheorem{lemma}{Lemma}
\newtheorem{theorem}{Theorem}

\newtheorem{definition}{Definition}

\title{
	\textbf{ \Large Towards an $O(\sqrt[3]{\log n})$-Approximation Algorithm for \\ {\sc Balanced Separator}}
}
\author{	\large{ Manjish Pal}\\\\
  \small{ Department of Computer Science and Engineering }\\
	\small{ Indian Institute of Technology Kanpur, INDIA. } \\
	\small{{\tt manjish@cse.iitk.ac.in}}\\
}
\date{}

\begin{document}

\pagestyle{empty}

\maketitle

\thispagestyle{empty}

\begin{abstract}
The {\sc $c$-Balanced Separator} problem is a graph-partitioning problem in which given a graph 
$G$, one aims to find a cut of minimum size such that both the sides of the cut
have at least $cn$ vertices. In this paper, we present new directions of progress in the {\sc $c$-Balanced Separator} problem. 
More specifically, we propose a new family of mathematical programs, 
which depends upon a parameter $\epsilon > 0$, and extend
the seminal work of Arora-Rao-Vazirani ({\sf ARV}) \cite{ARV} to show that the polynomial time solvability 
of the proposed family of programs implies an improvement in the approximation factor
to $O\left(\log^{\frac{1}{3} + \epsilon} n\right)$ from the best-known factor of $O(\sqrt{\log n})$ due to {\sf ARV}.
In fact, for $\epsilon = 1/3$, the program we get is the SDP proposed by {\sf ARV}.
For $\epsilon < 1/3$, this family of programs is not convex but one can transform them 
into so called \emph{\textbf{concave programs}} in which one optimizes a concave function over a convex feasible set. 
The properties of concave programs allows one to apply techniques due to Hoffman \cite{H81} or Tuy \emph{et al} \cite{TTT85} 
to solve such problems with arbitrary accuracy. But the problem of finding of a method to solve these programs 
that converges in polynomial time still remains open. Our result, although conditional, 
introduces a new family of programs which is more powerful than semi-definite programming 
in the context of approximation algorithms and hence it will of interest to investigate 
this family both in the direction of designing efficient algorithms and proving hardness results.
\end{abstract}

\newpage

\section{Introduction}
Graph partitioning is a problem of fundamental importance both in practice and theory. Many problems belonging 
to the several areas of computer science namely clustering, PRAM emulation, VLSI layout, packet routing in networks
can be modeled as partitioning a graph into two or more parts ensuring that the number of edges in the cut is ``small''. The word ``small'' doesn't refer to finding the min-cut in the graph as it doesn't ensure that the number of vertices in both sides of the cut is large. To enforce this balance condition one needs to normalize the cut-size in some sense. For the known notions of normalization like \emph{conductance}, \emph{expansion} and \emph{sparsity}, finding optimal separators is NP-hard for general graphs. Hence, the objective 
is to look for efficient approximation algorithms. 

Because of the huge amount of work done to design good approximation algorithm for these problems, 
graph partitioning has become one of the central objects of study in the theory of geometric 
embeddings and random walks. The first approximation algorithm for {\sc Graph Conductance}
came out of the study of the Reimannian Manifolds in form of the well known Cheegar's Inequality \cite{C70}
which says that if $\Phi(G)$ is the conductance of the graph and $\lambda$ is the second largest 
eigenvalue of graph Laplacian then $ 2\Phi(G) \geq \lambda \geq \Phi(G)^2/2$. 
Because of the quadratic factor in the lower bound, the true approximation is $\frac{1}{\Phi(G)}$ which 
in worst case can be $\Omega(n)$ in worst case. The first true approximation algorithm for {\sc Sparsest Cut}
and {\sc Graph Conductance} was designed by Leighton and Rao \cite{LR99} whose approximation factor was $O(\log n)$. 
This also gave an $O(\log n)$  \textbf{\emph{pseudo-approximation algorithm}} for {\sc $c$-Balanced Separator}. This
algorithm is referred to as a pseudo-approximation algorithm because 
instead of returning a $c$-balanced cut, it returns a $c'$-balanced cut for some fixed $c'<c$ whose 
expansion is at most $O(\log n)$ times the optimum expansion of best $c$-balanced cut. Their algorithm
was based on an LP framework motivated from the idea of Multi-commodity flows. Their main contribution
was to derive an approximate max-flow min-cut theorem corresponding to multi-commodity flow problem 
and the sparsest cut. Subsequently, a number of results
were discovered which showed that good approximation algorithms exist when one is considering
extreme cases such as the number of edges in the graphs is either very small or very large. In fact,
it is known for planar graphs one can find balanced cuts which are twice as optimal \cite{GSV94} and for 
graph with an average degree of $\Omega (n)$, one can design $(1+\epsilon)$-factor approximation algorithms
where $\epsilon > 0$ with running time polynomial in input size \cite{AKK} (such an algorithm is called a 
\emph{Polynomial Time Approximation Scheme} or PTAS). After 16 years the approximation factor
of $O(\log n)$ was improved to $O(\sqrt{\log n})$ in a breakthrough paper by Arora, Rao and Vazirani.
Their algorithm is based on semi-definite relaxations of these problems. 
The techniques and geometric structure theorems proved in their paper has subsequently led to breakthroughs
in the field of metric embeddings. The basic philosophy behind these approximation algorithms is to embed the
vertices of the input graph in an abstract space and derive a nice cut in this space. In the
linear programming approach one uses this abstract space as the $l_1$ metric \cite{LR99, LLR95, V02}. In the semi-definite 
programming framework used in \cite{ARV} one embeds the vertices on the surface of an $n$-dimensional unit sphere
such that they form an $l_2^2$ metric. The $l_2^2$ metric on the unit sphere translates into saying that for any three vectors
the angle subtended by any two among these at the third one is acute. 
One of the major tools used in this paper is the phenomenon of measure concentration on unit spheres. 

\textbf{Hardness Results:} Graph partitioning problems like {\sc Sparsest Cut} and {\sc Balanced Separator}
are considered to among the few NP-hard problems which have resisted various attempts to prove inapproximability 
results. After the result of {\sf ARV}, there has been 
a lot of impetus towards proving lower bounds on approximation factors. It has been 
shown by Ambuhl et al \cite{AMS07} that {\sc Sparsest Cut} can't have a PTAS unless NP-complete problems
can be solved in randomized sub-exponential time. Because of the strong connections 
between semi-definite programming and the {\sc Unique Games Conjecture (UGC)} of Khot \cite{Kh02}, certain inapproximability
results are also known which assume UGC. More specifically, Khot and Vishnoi \cite{KV05} show that 
UGC implies super-constant lower bounds on the approximation factor. In the following year,
Devanur et al \cite{DKSV06} showed that the integrality gap of the SDP relaxation of Arora-Rao-Vazirani is 
$\Omega(\log \log n)$ thereby disproving the original conjecture of ARV that 
the integrality gap of their SDP relaxation is atmost a constant. This result did not rely upon the {\sc UGC}. 
The recent progress towards designing efficient and good solutions to {\sc Unique Games}
has also been motivated from designing a reduction from {\sc Unique Games} to {\sc Sparsest Cut} \cite{AKK*08}. 

\subsection{Concave Programming}
In order to define \emph{\textbf{Concave Programming}} one first needs to define a concave
function. A concave function is the reverse of a convex function. 
Formally, a function $f:\mathbb{R}^d \rightarrow \mathbb{R}$ with domain \textbf{dom}$f$ is said to be concave if 
\textbf{dom}$f$ is convex and for all $x,y \in$ \textbf{dom}$f$, $f(\lambda x + (1-\lambda) y) \geq \lambda f(x) + (1-\lambda)f(y)$ for all $\lambda \in [0-1]$. Therefore, $f$ is concave iff $-f$ is a convex function. 
Based on this definition one defines concave programming as a form of mathematical programming 
in which one optimizes a concave function over a convex feasible set. 
More formally, a concave programming problem can be written as
$\displaystyle \left[ \min_{x \in C} f(x) \right ]$
where $C$ is a convex set in $\mathbb{R}^d$ and $f$ is a concave function. \\
Concave programming covers a broad range of 
non-linear global optimization problems which includes the well-known
DC(\emph{Difference of Convex Functions}) programming .
Due to its well-structured nature and wide applicability in economic problems as well as various other 
practical problems like allocation-location, water storage, standardization etc. \cite{TTT85},
there has been a lot of work in the field of optimization towards designing algorithms for various concave programming 
problems. One of the key properties of concave programming being exploited in these algorithms 
is a result that says that \emph{for every concave programming problem there is an extreme point
of the convex feasible set $C$ which globally minimizes the optimization problem}. 
The first algorithm for concave programming was designed by Tuy \cite{T64} in a restricted scenario
when the feasible set is a polytope. A more general case, when the feasible set is convex
but not necessarily polyhedral, was solved by Horst \cite{H76} and subsequently by Hoffman \cite{H81}, Tuy and
Thai \cite{TT81}. General concave programming is NP-hard as $\{0,1\}$-integer programming can
be cast as a concave program. There has been work towards designing efficient algorithms 
for some special class of concave programming. It has also been shown that some concave 
programs problems pertaining to Production-Transportation Problems can infact be solved in 
strongly polynomial time \cite{TGMV96}. A comprehensive list of works done in concave programming 
can be found in Vaserstein's homepage \cite{V}. \\
In this work, we introduce the use of a new family of concave programs towards designing
an improved approximation algorithm for the {\sc $c$-Balanced Separator} problem. 
  
\subsection {Contributions and Outline}
In section 2, we formally introduce the notions of sparsity and balanced cuts and
sketch the Semi-Definite relaxation for $c$-{\sc Balanced Separator}
of {\sf ARV}. We then start section 3 by introducing a family of relaxations for $c$-{\sc Balanced Separator} 
which is generated by a parameter $p > 0$. In section 4, using the techniques from \cite{ARV}, 
we show that one can improve the approximation factor to $O\left(\log ^{(1 + \frac{p}{2})/3} n\right)$ if
the proposed family of programs can be solved (by solving we mean getting a $(1+\epsilon)$-approximate answer) 
in polynomial time. Our result, although conditional, proposes new directions of progress on this problem 
and also a family of optimization problems which are more powerful than semi-definite programs 
in the context of approximation algorithms. Then in Section 5 show that one can transform this 
family of programs into a concave program, a form of mathematical programming in which one seeks to minimize 
a concave function over a convex feasible set. There are a number of algorithms which can solve such 
programs with arbitrary accuracy \cite{H81,TTT85}, although one is not guaranteed to
achieve a polynomial time convergence using these algorithms. 
Since this family is a new form of mathematical programming 
that is being used in an approximation algorithm, progress both in the direction of hardness and algorithms 
will provide more insights into the nature of these concave programs and can potentially lead us to 
optimal inapproximability results for various graph-partitioning problems. We end the paper with Section 6
in which we present conclusions and open problems.

\section {Preliminaries}
We now define the two versions of balanced graph partitioning problem namely the 
{\sc Sparsest Cut} and {\sc $c$-Balanced Separator}. It is well known that
upto constant factors approximating other versions of graph partitioning like
like {\sc Graph Conductance} and {\sc Uniform Sparsest Cut} are 
equivalent to approximating the {\sc Sparsest Cut}.
Although in this paper, we will mainly be concerned with the {\sc $c$-Balanced Separator} problem. 

\textbf{{\sc Sparsest Cut}} \\
Given a graph $G = (V,E)$ with $|V| = n, |E| = m$, for each cut $(S, \bar{S})$ define \emph{sparsity} of the cut to be the quantity $A(S) = \frac{|E(S, \bar{S})|}{|S|}$. The sparsest cut problem is to find $\alpha(G)$ where \\
$\displaystyle \alpha(G) = \min_{S \subset V, |S| < n/2} A(S)$. \\

\textbf{ $c$-{\sc Balanced Separator}} \footnote{In \cite{ARV} $c$-{\sc Balanced Separator} is defined as 
the minimum sparsity of $c$-balanced cuts, we will be working with a definition which upto constant factors
is equivalent to their definition} \\
Given a graph $G = (V,E)$ with $|V| = n, |E| = m$, the $c$-{\sc Balanced Separator} problem is to find $\alpha_c(G)$ where 
$\displaystyle \alpha_c(G) = \min_{S \subset V, cn < |S| < (1-c)n} E(S,\bar{S})$.

\subsection{SDP Relaxation for {\sc $c$-Balanced Separator}}
Unifying the spectral and the metric based (linear programming) approaches, {\sf ARV} used the following SDP
relaxation to get an improved (pseudo)-approximation algorithm
for the $c$-{\sc Balanced Separator}. Let us call this program $SDP_{BS}$,

\begin{center}
$\displaystyle \min \frac{1}{4}\sum_{i,j \in E} \|v_i - v_j\| ^2 $\\
$\displaystyle \|v_i\|^2 = 1 \quad \quad \forall i $\\
$\displaystyle \|v_i - v_j\|^2 + \|v_j - v_k\| ^2 \geq \|v_i - v_k\| ^2  \quad \quad \forall i,j,k$ \\
$\displaystyle \sum_{i<j} \|v_i - v_j\| ^2 \geq 4c(1-c)n^2$
\end{center}

It is easy to see that this indeed is a \emph{vector program} (and hence an SDP)
and is a relaxation for the $c$-{\sc Balanced Separator} 
problem. To show that this is a relaxation we have to show that for every cut we can get an assignment of vectors
such that all the constraints are satisfied and the value of the objective function
is the size of the cut. Given a cut $(S,\bar{S})$ if one maps all the vertices in 
$S$ to a unit vector $\textbf{n}$ and the vertices in $\bar{S}$ to $-\textbf{n}$ then the value
of the function is indeed the cardinality of $E(S,\bar{S})$. 
The main idea behind their algorithm is to show that for any set of vectors which satisfy the constraints
of the SDP there always exist two disjoint subsets of ``large'' size such that for any two points belonging to 
different subsets the squared Euclidean distance between them is atleast $\Omega\left(\frac{1}{\sqrt{\log n}}\right)$.
The same idea is also used to get an improved approximation algorithm for {\sc Sparsest Cut} in \cite{ARV}.
Subsequently, this key idea has crucially been used in various other SDP based approximation algorithms and in 
solving problems related to metric embeddings.  

\section{A New Relaxation for $c$-{\sc Balanced Separator}}
Consider the following family of optimization problems which depend on a parameter $p \geq 0$.
This family is essentially an extension of the semi-definite program proposed by {\sf ARV}.
Throughout the paper we will use $\|.\|$ to represent the $l_2$ norm. Let us call this 
family of programs $F_{BS}^p$.

\begin{center}
$\displaystyle \min \frac{1}{2^p}\sum_{i,j \in E} \|v_i - v_j\| ^p $\\
$\displaystyle \|v_i\|^2 = 1 \quad \quad \forall i $\\
$\displaystyle \|v_i - v_j\|^p + \|v_j - v_k\| ^p \geq \|v_i - v_k\| ^p  \quad \quad \forall i,j,k$ \\
$\displaystyle \sum_{i,j \in E} \|v_i - v_j\| ^2 \geq 4c(1-c)n^2$
\end{center}

Note that for $p = 2$ this is the SDP relaxation proposed by {\sf ARV}. For $p = 1$,               
we are mapping the points onto a unit sphere, therefore
we do not have to force the additional triangle inequality constraint of $l_2$ metric. 
The same mapping described for $SDP_{BS}$ of the vertices of the graph
onto the unit sphere allows us to conclude that each program in this family 
is also a relaxation for {\sc $c$-Balanced Separator}.
We will show that the techniques used in \cite{ARV} for lower bounding the optimum
value of their semi-definite program can be extended in this case as well by
appropriately modifying the ingredients of Theorem 1 of their paper. Under the assumption that we can solve $F_{BS}^p$ in polynomial time
for any $p$, $0<p<2$, we are able achieve an approximation factor of $O\left(\log^{\frac{1}{3} + \frac{p}{6}} n\right)$. 
We first show how to modify the results of \cite{ARV}, thereby reducing the problem of obtaining 
an improved approximation algorithm to that of finding a polynomial time algorithm for solving 
the family of programs mentioned above.

\section{ARV Proof Modifications}
We first modify the definition of $\Delta$-separated sets 
\begin{definition}[$\Delta$-{\sc Separated Sets}]
Two sets of vectors in $\mathbb{R}^d$, $S$ and $T$ are said to be $\Delta$-separated if for all $v \in S$ and $u \in T$
$\|v - u\|^p \geq \Delta$. 
\end{definition}

\begin{definition}
For $p > 0$, a set of vectors in $\mathbb{R}^d$ is said to be a unit $c$-spread $l_2^p$ representation if they
satisfy the last three constraints in the program $F^{p}_{BS}$.
\end{definition}

Under the new definition of $\Delta$-separated sets the main theorem of {\sf ARV} can 
be modified in the following way:
\begin{theorem}\label{separated}
For every $c' > 0$, there are constants $c, b > 0$ such that every $c$-spread unit-$l_2^p$
representation with $n$ points contains $\Delta$-separated subsets $S, T$ of size $c'n$, 
where $\Delta = b \log ^{-(1 + \frac{p}{2})/3} n$. Also, there is a randomized polynomial-time
algorithm for finding these subsets $S, T$.
\end{theorem}

This theorem immediately allows us to conclude the following result.
\begin{theorem}
Given a graph $G = (V,E)$, if the program $F^{p}_{BS}$ can be solved in polynomial time for a fixed $p$, then there 
exists a randomized $\displaystyle O\left(\log ^{(1 + \frac{p}{2})/3} n\right)$-pseudo approximation algorithm for 
$c$-{\sc Balanced Separator}. 
\end{theorem}

\begin{proof}\emph{(Sketch)}
For a fixed $p$, let $U = \{u_1,u_2, \dots, u_n\}$ is the optimum solution to the program $F^{p}_{BS}$ with the optimum value
value as $\cal O$. We construct the weighted graph $G'$ on the vertex set of the original graph, 
s.t. for every edge $(v_i, v_j)$ 
we impose a weight of $\frac{1}{2^p} \|v_i - v_j\|^p$. Now apply the algorithm of Theorem \ref{main}
to obtain two subsets $S',T'$ of $U$ of size at least $c'n$ which are 
$\displaystyle \Delta = b \log ^{-(1 + \frac{p}{2})/3} n$-separated. 
Let $S,T$ be the corresponding sets of vertices in $V$. Pick a number $r$ 
randomly uniformly from the range [0-$\Delta$] and report the cut $(V_r, \bar{V_r})$
as the answer where $V_r$ is the set of all vertices within distance $r$ from $S$. 
Performing a similar analysis as in Corollary 2 of \cite{ARV}, it is easy to show that with high 
probability $(V_r, \bar{V_r})$ is a $\displaystyle O(\log ^{(1 + \frac{p}{2})/3} n )$-approximate 
$c'$-balanced cut.
\end{proof}

\subsection{Proof of Theorem \ref{separated}}

\begin{algorithm}[ht]
\caption{\textbf{{\sc Modified}-Set-Find}}
\KwIn{A set of vectors $V = \{v_1,v_2, \dots v_n\}$ in $\mathbb{R}^d$
which form a unit $c$-spread $l_2^p$ representation and parameters $\Delta$ and $\sigma$.}
\KwOut{Two sets $S$ and $T$ which are $\Delta$-separated with the desired balance $c'$.}
\begin{itemize}
\item[1.] Pick a random unit vector $u$. 
\item[2.] Let $v_k$ be the vector that realizes the median of the values taken by 
$\left \langle v_i,u \right \rangle$ for \\
$i = 1,2 \dots n$ and let $m$ be the median value. 
\item[3.] Let $S'$ be the set of vectors in $V$ satisfying $\left \langle v_i,u \right \rangle \geq  m + \frac{\sigma}{2\sqrt{d}}$
and $T'$ be the \\
vectors in $V$ which satisfy $\left \langle v_i,u \right \rangle \leq m - \frac{\sigma}{2\sqrt{d}}$.
\item[4.] If $|S'| \leq 2c'n$ or $|T'| \leq 2c'n$, HALT. Otherwise remove all the vectors  
$v_{i_1} \in S'$ \\
and $v_{i_2} \in T'$ which satisfy $\|v_{i_1} - v_{i_2}\|^p \leq \Delta$.
\end{itemize}
\textbf{Return}: Remaining sets as $S$ and $T$.
\end{algorithm}

Given a unit $c$-spread $l_2^p$ representation of vectors, 
the algorithm to find two $\Delta = b \log ^{-(1 + \frac{p}{2})/3} n$-separated sets needs
a small modification over the Set-Find algorithm of {\sf ARV} which is presented as 
{\sc Modified}-\textbf{Set-Find}. Notice that the modification is made at the last step
when the algorithm is discarding pairs.

In order to prove our claim for {\sc Modified}-\textbf{Set-Find} 
we will borrow the definitions of $(\sigma, \delta, c')$-matching cover, 
$(\sigma,\delta)$-uniform matching cover and $(\epsilon, \delta)$-cover directly from \cite{ARV}. Among these
we will only reproduce the definitions of $(\sigma, \delta, c')$-matching cover and $(\epsilon, \delta)$-cover.
The basic idea is that the all these notions of covers do not depend on the triangle inequality of $l_2^2$ metric and
hence they also make sense for $l_2^{p}$ representations. 

\begin{definition}
For a set $V$ of vectors, a $(\sigma,\delta,c')$-matching cover is a set of (partial) matchings $\cal M$
such that for at least $\delta$ fraction of directions $u$ there exists a matching $M_u \in $ $\cal M$
with size at least $c'n$ such that for each pair $(v_i,v_j)$ in the matching 
$\left \langle v_i-v_j,u \right \rangle \geq 2\sigma/\sqrt{d}$.
\end{definition}

Let $M$ be the multi-graph obtained by the union of all the matchings in $\cal M$.

\begin{definition}
A set of vectors $w_1,w_2,\dots, w_n$ is is said to be an $(\epsilon, \delta)$-cover if $\|w_i\| \leq 1$ for all $i$
and for at least $\delta$ fraction of the directions there exist  $i \in [n]$ 
$\left \langle w_i,u \right \rangle \geq \epsilon$. A set of vectors if said to $(\epsilon,\delta)$-cover 
a point $x$, if the set of vectors $\{x-w_i | i \in [n]\}$ is a $(\epsilon, \delta)$-cover.
\end{definition}

The most important thing to note is that the well-separated constraint 
($\sum_{i,j \in E} \|v_i - v_j\| ^2 \geq 4c(1-c)n^2$)
is the same for both $SDP_{BS}$ and $F_{BS}^{p}$. This
allows us to conclude that Lemmas 3-7 of \cite{ARV} all hold for any set 
of unit $c$-spread $l_2^{p}$ representations  as well. Only Theorem 8 of \cite{ARV}
needs considerable changes which we present as Theorem \ref{main}. For the sake of 
completeness we will reproduce the necessary ingredients used in the proof of 
Theorem 8 of \cite{ARV} namely the definition of $k$-core and Lemma 7. 
But before going into the proof of our version of Theorem 8, let us recall the behavior of 
projection of a random unit vector onto a fixed vector.

\begin{lemma} \label{gaussian}
If $v$ is a vector of length $l$ in $\mathbb{R}^d$ and $u$ is a randomly chosen unit vector 
\begin{itemize}
\item for $x < 1$, $\Pr\{|\left \langle v,u \right \rangle |\leq \frac{xl}{\sqrt{d}}\} \leq 3x$.
\item for $x \leq \frac{\sqrt{d}}{4}$, $\Pr\{|\left \langle v,u \right \rangle| \geq \frac{xl}{\sqrt{d}}\} \leq e^{-x^2/4}$.
\end{itemize}
\end{lemma}

\begin{definition}
Given a set of $n$ points about that is $(\sigma, \delta, c')$ matching covered by $\cal M$ with 
associated matching graph $M$, define $v$ to be in the $k$-core, $S_k$ if $v$ is $(\frac{k\sigma}{2\sqrt{d}},\frac{1}{2})$-covered
by points which are within $k$ hops of $v$ in the matching graph.
\end{definition}

The following lemma (Lemma 7 of {\sf ARV}) captures an important property of matching covers.
A crucial result used in its proof is Levy's iso-perimetric inequality and measure concentration
on spheres \cite{B97, Mat02}.

\begin{lemma}\label{metric-lemma}
For every set of $n$ points that is $(\sigma, \delta, c')$-matching covered by $\cal M$ with associated 
matching graph $M$, there are positive constants $a=a(\delta,c')$ and $b = b(\delta,c')$ such that for 
every $k \geq 1$ one of the following holds: 
\begin{itemize}
\item[1.] $|S_k| \geq a^kn$.
\item[2.] There is a pair with distance at most $k$ in the matching graph $M$ such that $\|v_i-v_j\| \geq \frac{b\sigma}{\sqrt{k}}$.
\end{itemize}
\end{lemma}

The following lemma can be used to prove the main theorem which shows that the algorithm
{\sc Modified}-\textbf{Set-Find} succeeds with constant probability. 

\begin{theorem}\label{main}
The \mbox{{\sc Modified}-\textbf{Set-Find}} algorithm finds a $\Delta$-separated set for a $c$-spread $l_2^p$ representation
with constant probability for $\Delta = \Theta(\log^{\beta} n)$ where $\beta = \frac{1 + \frac{p}{2}}{3}$.
\end{theorem}
\begin{proof}
If {\sc Modified}-\textbf{Set-Find} fails with probability $1-\delta$,
then according to the definition of matching covers it can be shown that,
the set of deleted points will be $(\sigma, \delta/2, c')$-matching covered. Let $M$ be the 
associated matching graph. This implies that 
we can use Lemma \ref{metric-lemma} for the set of points. We will show that 
in such a situation both the cases of Lemma \ref{metric-lemma}
do not hold which in turn implies that {\sc Modified}-\textbf{Set-Find} does not fail with high probability. 
We first start with dispensing the case 2 of Lemma \ref{metric-lemma}. Since the points lie in a $l_2^p$ metric
and $v_i$ and $v_j$ are within $k$-hops in the corresponding matching graph $M$ we will have
$\|v_i - v_j\|^p \leq k\Delta$ which implies $\|v_i - v_j\| \leq \sqrt[p]{k\Delta}$. Now let us 
choose $k = \frac{(b\sigma/2)^{\alpha}}{\Delta ^r}$ where $r = \frac{1}{1 + \frac{p}{2}}$ and 
$\alpha = \frac{1}{(\frac{1}{2} + \frac{1}{p})}$. Under this choice of $k$, 
one can verify that $\frac{b\sigma}{\sqrt{k}} > \sqrt[p]{k\Delta}$ 
which will lead us to a contradiction. 

Now we consider the case 1 of Lemma \ref{metric-lemma}. This says that the number of vectors
which are $(\frac{k \sigma}{\sqrt{d}}, \frac{1}{2})$ covered by points within $k$-hops
of the matching graph $M$, is at least some constant fraction of the total number of points. Consider 
a point $v_i$ which is in the $k$-core as defined above. Let $v_j$ be a point that belongs to the 
set that $(\frac{k \sigma}{\sqrt{d}}, \frac{1}{2})$-covers $v_i$. Now by definition of $k$-core  
with at least probability $\frac{1}{2}$, $\left \langle v_j-v_i,u \right \rangle \geq \frac{k\sigma}{2\sqrt{d}}$.
But because of the fact that the points come from a $l_2^p$ metric with in $k$ hops 
$\|v_i - v_j\| \leq \sqrt[p]{k\Delta}$. Now using the Gaussian behavior of projections for a 
vector $v_i-v_j$ a randomly chosen unit vector $u$ satisfies, 
$\textbf{Pr}\displaystyle \left \{\left \langle v_j-v_i,u \right \rangle \geq \frac{x \|v_i - v_j\|}{\sqrt{d}}\right \} \leq e^{-\frac{x^2}{4}}$. \\
Therefore taking $x = \frac{k \sigma}{2 \|v_i - v_j\|}$ we get,
$\textbf{Pr}\displaystyle \left \{\left \langle v_j-v_i,u \right \rangle \geq \frac{k\sigma \|v_i - v_j\|}{2 \|v_i - v_j\|\sqrt{d}}\right \} 
\leq e^{-\frac{k^2 \sigma^2}{16 \|v_i - v_j\|^2}} 
\leq e^{-\frac{b^{2\alpha} {\sigma}^{2\alpha} \sigma^2}{16 \Delta^{2r} (k\Delta)^\frac{2}{p}}} $
If we denote the exponent of $e$ by $A$, then we have
$\displaystyle A = -\frac{b^{2\alpha - \frac{2\alpha}{p}} {\sigma}^{2 + 2 \alpha -  \frac{2\alpha}{p}} 2^{\frac{2\alpha}{p}}}{16 \Delta^{2r} \cdot \Delta^{\frac{2(1-r)}{p}}}$.
Since $\alpha = \frac{1}{(\frac{1}{2} + \frac{1}{p})}$, we have the exponent of $\sigma$ as
$\displaystyle \left(2 + 2 \alpha -  \frac{2\alpha}{p} \right) = 2 - \frac{2}{(\frac{p}{2} + 1)} + 2 \alpha = 2 \alpha + \frac{2p}{p + 2} > 0$.
Let $\displaystyle \gamma = \frac{b^{2\alpha - \frac{2\alpha}{p}} {\sigma}^{2\alpha + \frac{2p}{p + 2}} 4^{\frac{1}{1+p/2}}}{16} $.
Now one can choose a small constant $g$ such that $\Delta = \displaystyle \frac{g}{\log^{\beta} n}$ for $\beta < 1$ and $\displaystyle \frac{\gamma}{g^{2r + \frac{2(1-r)}{p}}} > 4$, 
which can be done because for a fixed $p$ $\gamma$ is a constant and we are going to set $2\beta (r + \frac{1-r}{p}) = 1$.
We therefore get the desired probability to be atmost
$\displaystyle e^{-\gamma(\log n)^{2\beta (r + \frac{1-r}{p})}}$. Putting $r = \frac{1}{1+ \frac{p}{2}}$ we get the exponent of $(\log n)$ as 
$\displaystyle 2\beta\left( \frac{1}{1+\frac{p}{2}} + \frac{1- \frac{1}{1 + p/2}}{p}\right) = \beta \left(\frac{3}{1 + \frac{p}{2}}\right)$.
Therefore, if we choose $\beta = \displaystyle \frac{1 + \frac{p}{2}}{3}$, we can get this probability as atmost $\frac{1}{n^4}$.
Now, clearly the probability that a vector $v_i$ is covered by points within $k$-hops of matching graph is atmost the probability
that there exists two points $v_i$ and $v_j$ such that for a random unit vector $u$, the above event occurs.
From the above calculation, the probability that such an event occurs for any pair $v_i,v_j$ is 
less that $O(\frac{1}{n^2})$ via the union bound contradicting the condition of Lemma \ref{metric-lemma} that this probability is at least 1/2.
\end{proof}

\textbf{Some Discussion on the Result}: \\
It is not clear whether this method will receive 
benefits from the stronger version of Lemma \ref{metric-lemma} of \cite{ARV} in which they prove that 
for the second case $\|v_i - v_j\| \geq \sigma$ and use it along with other ideas
so that their algorithm works even for $\Delta = \left(\frac{1}{\sqrt{\log n}}\right)$. The main
reason is that if we try to take a $k$ that is of the form chosen in Theorem \ref{main}
then we don't get a dependence of $r$ in terms of $p$ and therefore we don't get a parametrization
of the approximation factor in terms of $p$. Although one might come up with a method such that the above mentioned result 
can also be used to get an improvement over this bound of $\Delta$. One can also ask the 
question, why did we choose to set $ \beta \left(\frac{3}{1 + \frac{p}{2}}\right)$ as 1
because we could have improved the bound on probability if we had chosen a value greater than 1
but in that case one can easily notice that we would have to sacrifice with the approximation factor
and we would have got value a value of $\beta$ which is worse than this value.

\section{A Concave Programming Formulation}
In this section, we consider the family of optimization problems $F_{BS}^{p}$ proposed above 
and transform it into a concave program. This formulation allows us to use the algorithms which have been developed
to solve a concave program with arbitrary accuracy. 
We now write $F_{BS}^{p}$ as a program with variables as matrix entries and not as $d$-dimensional
vectors. The variables in the new program are of the form $x_{ij} = \left \langle v_i,v_j \right \rangle$. Since all $v_i$'s
are unit vectors we can write $\|v_i - v_j\|$ as $\sqrt{2 - 2\left \langle v_i,v_j \right \rangle}$.
If we consider the matrix $X$ with $ij^{th}$ entry as $x_{ij}$ we can write the above problem as

\begin{center}
$\displaystyle \min \frac{1}{2^{p/2}} \sum_{i,j \in E} (1 - x_{ij})^{p/2} $\\
$\displaystyle x_{ii} = 1 \quad \quad \forall i $\\
$\displaystyle (1 - x_{ij})^{p/2} + (1 - x_{jk})^{p/2} \geq (1 - x_{ik})^{p/2}  \quad \quad \forall i,j,k$ \\
$\displaystyle \sum_{i < j } (1 - x_{ij}) \geq c(1-c)n^2$ \\
$\displaystyle X \succeq 0$
\end{center}
where $X \succeq 0$ means $X$ is positive semi-definite. 

As we have seen earlier that in order to get an improved approximation factor 
we must have $p < 1$. Under such a restriction the problem becomes a non-convex 
feasibility problem as the function $(1 - x_{ij})^{p/2}$ is not convex. This 
is a crucial deviation from all the relaxations which have been studied till
now in the context of approximation algorithms. 
Because of the non-convex nature of the problem we can't use any of the 
well known techniques like the ellipsoid method and the interior point methods
and hence can't directly guarantee the polynomial time solvability of the program.
We therefore transform it into a form which allows us to prove some interesting properties. 
In the above program if we do a change of variable,  $z_{ij} = (1 - x_{ij})$ for all $i,j = 1,2 \dots n$,
the minimization problem looks as the following:

\begin{center}
$\displaystyle \min \frac{1}{2^{p/2}} \sum_{i,j \in E} z_{ij}^{p/2} $\\
$\displaystyle z_{ij}^{p/2} + z_{jk}^{p/2} \geq z_{ik}^{p/2}  \quad \quad \forall i,j,k$ \\
$\displaystyle \sum_{i < j} z_{ij} \geq c(1-c)n^2$ \\
$\displaystyle z_{ii} = 0 \quad \quad \forall i $\\
$\displaystyle \textbf{1} - Z \succeq 0$
\end{center}
where \textbf{1} is the matrix with all entries as 1. \\
Let us call the above program $\tilde{F}_{BS}^{p}$.
This formulation allows us to prove the following lemma:

\begin{lemma}\label{concave}
$\tilde{F}_{BS}^{p}$ is a concave program for $0<p<2$. 
\end{lemma}
\begin{proof}
Since $z^{p/2}$ is concave for $p<2$ for $z>0$, and the sum of concave functions
is also concave, the objective function is clearly concave. For the constraints
defining the feasible set, $\sum_{i < j} z_{ij} \geq c(1-c)n^2$ and $z_{ii} = 0$
are convex. The constraint $\textbf{1} - Z \succeq 0 $ can be shown to be convex 
as follows: Let $Z_1$ and $Z_2$ be two matrices corresponding to the variables $z_{ij}$'s
which lie in the feasible set. Therefore, they satisfy $\textbf{1} - Z_1 \succeq 0$ and 
$\textbf{1} - Z_2 \succeq 0$. Now, consider the line segment for $\lambda \in [0-1]$
$\lambda Z_1 + (1-\lambda) Z_2$ and the matrix $\textbf{1}-(\lambda Z_1 + (1-\lambda) Z_2)$. This
is positive semidefinite as it can be rewritten as $\lambda(\textbf{1}- Z_1) + (1-\lambda)(\textbf{1} - Z_2)$ 
which is a sum of two PSD matrices. \\
The only type of constraint left are the triangle inequality constraints. Consider 
an inequality of this type say $z_{ij}^{p/2} + z_{jk}^{p/2} \geq z_{ik}^{p/2}$.  
In general, let us look at the region $x^{r} + y^{r} \geq z^r$ for $0 < r < 1$.
If $r = 1/q$ for $q > 1$ then this region is same as $ \left(x^{1/q} + y^{1/q}\right)^{q} \geq z$.
Let $p_1 = (x_1,y_1,z_1)$ and $p_2 = (x_2,y_2,z_2)$ be two points which lie in this region, i.e. 
$\left({x_1}^{1/q} + {y_1}^{1/q}\right)^{q} \geq z_1$ and 
$\left({x_2}^{1/q} + {y_2}^{1/q}\right)^{q} \geq z_2$. To prove 
the convexity of the region we need to show that for any $\lambda \in [0-1]$,
$(\lambda x_1 + (1-\lambda) x_2, \lambda y_1 + (1-\lambda) y_2, \lambda z_1 + (1-\lambda) z_2)$
also lies inside the region for all such points $p_1$ and $p_2$. Therefore, we have to show
$\displaystyle \lambda z_1 + (1-\lambda) z_2 \leq \left((\lambda x_1 + (1-\lambda) x_2)^{\frac{1}{q}} + (\lambda y_1 + (1-\lambda) y_2)^\frac{1}{q} \right)^{q}$.
Thus we will be done if we show 
$\displaystyle \lambda ({x_1}^{1/q} + {y_1}^{1/q})^{q} + (1-\lambda) ({x_2}^{1/q} + {y_2}^{1/q})^{q} \leq \left((\lambda x_1 + (1-\lambda) x_2)^{\frac{1}{q}} + (\lambda y_1 + (1-\lambda) y_2)^\frac{1}{q} \right)^{q}$.
which is equivalent to proving that the function $f(x,y) = \left({x}^{\frac{1}{q}} + {y}^{\frac{1}{q}}\right)^{q}$
is concave. We will prove this by showing that the Hessian of this function is negative-definite for all $x,y$. We now
compute the entries of the Hessian matrix. The following calculations are easy to verify,
\begin{eqnarray*}
\frac{\partial f}{\partial x} &=& \left(1 + \frac{y^{\frac{1}{q}}}{x^{\frac{1}{q}}} \right)^{q-1} ; 
\frac{\partial f}{\partial y} = \left(1 + \frac{x^{\frac{1}{q}}}{y^{\frac{1}{q}}} \right)^{q-1} ; 
\frac{\partial^2 f}{\partial x^2} = -\left(\frac{q-1}{q}\right) \left(1 + \frac{x^{\frac{1}{q}}}{y^{\frac{1}{q}}} \right)^{q-2} \frac{y^{\frac{1}{q}}}{x^{\frac{q+1}{q}}} ;\\
\frac{\partial^2 f}{\partial y^2} &=& -\left(\frac{q-1}{q}\right) \left(1 + \frac{y^{\frac{1}{q}}}{x^{\frac{1}{q}}} \right)^{q-2}  \frac{x^{\frac{1}{q}}}{y^{\frac{q+1}{q}}} ;  \frac{\partial^2 f}{\partial x \partial y} = \left(\frac{q-1}{q}\right) \left( \frac{1}{x^{\frac{1}{q}}} + \frac{1}{y^{\frac{1}{q}}}\right)^{q-2} \frac{1}{y^{\frac{1}{q}}x^{\frac{1}{q}}} = \frac{\partial^2 f}{\partial y \partial x}
\end{eqnarray*}

In order to show that the Hessian is negative-definite we have to show that for any $\alpha, \beta \in \mathbb{R}$,
the following expression is always non-positive for all $x,y > 0$ (for $x,y$ as 0 the derivatives do not exist):
\begin{eqnarray*}
&& \alpha^2 \frac{\partial^2 f}{\partial x^2} + \beta^2 \frac{\partial^2 f}{\partial y^2} +  2 \alpha \beta \frac{\partial^2 f}{\partial x \partial y} \\
& = & -\left( \frac{q-1}{q} \right) \left[ \alpha^2 \left(1 + \frac{x^{\frac{1}{q}}}{y^{\frac{1}{q}}} \right)^{q-2}  \frac{y^{\frac{1}{q}}}{x^{\frac{q+1}{q}}} + \beta ^2 \left(1 + \frac{y^{\frac{1}{q}}}{x^{\frac{1}{q}}} \right)^{q-2} \frac{x^{\frac{1}{q}}}{y^{\frac{q+1}{q}}} - \left( \frac{1}{x^{\frac{1}{q}}} + \frac{1}{y^{\frac{1}{q}}}\right)^{q-2} \cdot \frac{2\alpha \beta}{y^{\frac{1}{q}}x^{\frac{1}{q}}}\right] \\
&=& -\left( \frac{q-1}{q} \right) \left({x}^{\frac{1}{q}} + {y}^{\frac{1}{q}}\right)^{q} \left[ \frac{\alpha^2y^{\frac{1}{p}}}{x^{\frac{2q-1}{q}}} + \frac{\beta^2x^{\frac{1}{q}}}{y^{\frac{2q-1}{q}}} - \frac{2\alpha \beta}{x^{\frac{q-1}{q}}y^{\frac{q-1}{q}}} \right] \\
& = & -\left( \frac{q-1}{q} \right) \left({x}^{\frac{1}{q}} + {y}^{\frac{1}{q}}\right)^{q} \left[ 
\frac{\alpha^2y^2 + \beta^2x^2 - 2\alpha \beta xy }{x^{\frac{2q-1}{q}}y^{\frac{2q-1}{q}}} \right] 
= -\left( \frac{q-1}{q} \right) \left({x}^{\frac{1}{q}} + {y}^{\frac{1}{q}}\right)^{q} \left[ 
\frac{(\alpha y - \beta x)^2 }{x^{\frac{2q-1}{q}}y^{\frac{2q-1}{q}}} \right] 
\end{eqnarray*}
which is non-positive for all $\alpha, \beta$
This proves that the region $x^{p/2} + y^{p/2} \geq z^{p/2}$ is a convex set for all $0< p < 2$. Hence the 
intersection of all the triangle inequality constraints is also convex.
\end{proof}

\section{Conclusion}
In this paper, we introduced a new family of mathematical programs inspired from the
well-known semi-definite program of {\sf ARV} that promises a 
$\displaystyle O\left (\log^{\frac{1}{3} + \epsilon} n\right)$ pseudo-approximation 
algorithm for $c$-{\sc Balanced Separator} under the condition that the family 
of programs can be solved in polynomial time. Since this family provably gives better 
approximation guarantees than the celebrated linear and semi-definite relaxations
of Leighton-Rao \cite{LR99} and Arora-Rao-Vazirani \cite{ARV} respectively, 
investigation both in the direction of polynomial time solvability 
or hardness will be highly interesting. The formulation of the proposed family of programs
into a well-structured form of mathematical programming called concave programming
also gives us hope that for many of the problems for which optimal approximation factors
are not known one can possibly rely upon some ``nice'' programs 
which are although not convex but can be potential candidates for 
polynomial time solvability. Given that some algorithms for 
solving concave programs are easy simple to comprehend, it would be also 
interesting to know whether one can analyze their runs on $F_{BS}^{p}$ 
and prove polynomial time convergence. Another area of investigation could be investigating the  
links of this family of programs with the {\sc UGC} 
which has been able to show optimality (close to optimality) of various approximation algorithms based on SDP. \\

\section{Acknowledgments}
I would like to thank Prof. Sumit Ganguly and Prof. Shashank K. Mehta
for having some stimulating discussions on the problem. Thanks to Purushottam 
Kar for having discussions at various points during the work.

\end{document}